\documentstyle[epsf,12pt]{article}

\makeatletter

 \@addtoreset{equation}{section}
\makeatother

\begin{document}
\topmargin -35pt
\oddsidemargin 5mm

%%%%%%  user's commands  %%%%%%%%%%%%%%%%%%%%%%%%%%%%%%%%%%%%%%%%%%%
\newcommand {\beq}{\begin{eqnarray}}
\newcommand {\eeq}{\end{eqnarray}}
\newcommand {\non}{\nonumber\\}
\newcommand {\eq}[1]{\label {eq.#1}}
\newcommand {\defeq}{\stackrel{\rm def}{=}}
\newcommand {\gto}{\stackrel{g}{\to}}
\newcommand {\hto}{\stackrel{h}{\to}}
\newcommand {\1}[1]{\frac{1}{#1}}
\newcommand {\2}[1]{\frac{i}{#1}}
\newcommand {\th}{\theta}
\newcommand {\thb}{\bar{\theta}}
\newcommand {\ps}{\psi}
\newcommand {\psb}{\bar{\psi}}
\newcommand {\ph}{\varphi}
\newcommand {\phs}[1]{\varphi^{*#1}}
\newcommand {\sig}{\sigma}
\newcommand {\sigb}{\bar{\sigma}}
\newcommand {\Ph}{\Phi}
\newcommand {\Phd}{\Phi^{\dagger}}
\newcommand {\Sig}{\Sigma}
\newcommand {\Phm}{{\mit\Phi}}
\newcommand {\eps}{\varepsilon}
\newcommand {\del}{\partial}
\newcommand {\dagg}{^{\dagger}}
\newcommand {\pri}{^{\prime}}
\newcommand {\prip}{^{\prime\prime}}
\newcommand {\pripp}{^{\prime\prime\prime}}
\newcommand {\prippp}{^{\prime\prime\prime\prime}}
\newcommand {\pripppp}{^{\prime\prime\prime\prime\prime}}
\newcommand {\delb}{\bar{\partial}}
\newcommand {\zb}{\bar{z}}
\newcommand {\mub}{\bar{\mu}}
\newcommand {\nub}{\bar{\nu}}
\newcommand {\lam}{\lambda}
\newcommand {\lamb}{\bar{\lambda}}
\newcommand {\kap}{\kappa}
\newcommand {\kapb}{\bar{\kappa}}
\newcommand {\xib}{\bar{\xi}}
\newcommand {\ep}{\epsilon}
\newcommand {\epb}{\bar{\epsilon}}
\newcommand {\Ga}{\Gamma}
\newcommand {\rhob}{\bar{\rho}}
\newcommand {\etab}{\bar{\eta}}
\newcommand {\chib}{\bar{\chi}}
\newcommand {\tht}{\tilde{\th}}
\newcommand {\zbasis}[1]{\del/\del z^{#1}}
\newcommand {\zbbasis}[1]{\del/\del \bar{z}^{#1}}
\newcommand {\vecv}{\vec{v}^{\, \prime}}
\newcommand {\vecvd}{\vec{v}^{\, \prime \dagger}}
\newcommand {\vecvs}{\vec{v}^{\, \prime *}}
\newcommand {\alpht}{\tilde{\alpha}}
\newcommand {\xipd}{\xi^{\prime\dagger}}
\newcommand {\pris}{^{\prime *}}
\newcommand {\prid}{^{\prime \dagger}}
\newcommand {\Jto}{\stackrel{J}{\to}}
\newcommand {\vprid}{v^{\prime 2}}
\newcommand {\vpriq}{v^{\prime 4}}
\newcommand {\vt}{\tilde{v}}
\newcommand {\vecvt}{\vec{\tilde{v}}}
\newcommand {\vecpht}{\vec{\tilde{\phi}}}
\newcommand {\pht}{\tilde{\phi}}
\newcommand {\goto}{\stackrel{g_0}{\to}}
\newcommand {\tr}{{\rm tr}\,}
\newcommand {\GC}{G^{\bf C}}
\newcommand {\HC}{H^{\bf C}}
\newcommand{\vs}[1]{\vspace{#1 mm}}
\newcommand{\hs}[1]{\hspace{#1 mm}}

%%%%%%%%%%%%%%%%%%%%%%%%%%%%%%%%%%%%%%%%%%%%%%%%%%%%
\setcounter{page}{0}

%%%%%%%%% title %%%%%%%%%%%%%
\begin{titlepage}

\begin{flushright}
OU-HET 349\\
TIT/HEP-450\\
hep-th/0006027\\
June 2000
\end{flushright}
\bigskip

\begin{center}
{\LARGE\bf
K\"{a}hler Normal Coordinate Expansion in Supersymmetric Theories
}
\vs{10}

\bigskip
{\renewcommand{\thefootnote}{\fnsymbol{footnote}}
{\large\bf Kiyoshi Higashijima$^a$\footnote{
     E-mail: higashij@phys.sci.osaka-u.ac.jp.}
 and Muneto Nitta$^b$\footnote{
E-mail: nitta@th.phys.titech.ac.jp.}
}}

\setcounter{footnote}{0}
\bigskip

{\small \it
$^a$Department of Physics,
Graduate School of Science, Osaka University,\\
Toyonaka, Osaka 560-0043, Japan\\
$^b$Department of Physics, Tokyo Institute of Technology,\\ 
Oh-okayama, Meguro, Tokyo 152-8551, Japan\\
}

\end{center}
\bigskip

%%%%%%%%% abstract %%%%%%%%
\begin{abstract}
The Riemann normal coordinate expansion method is 
generalized to a K\"ahler manifold. 
The K\"ahler potential 
and holomorphic coordinate transformations are used 
to define normal coordinates preserving the complex structure.
The existence of these K\"ahler normal coordinate is shown explicitly 
to all orders. The formalism is applied to background field 
methods in supersymmetric nonlinear sigma models. 
\end{abstract}

\end{titlepage}

%%%%%%%%%%%%%%%%%%%%
\section{Introduction}
In nonlinear sigma models, a field variable $\varphi(x)$, defined 
at a space-time point $x$, takes a value on a Riemannian manifold 
called the target manifold.
Its $S$-matrices are invariant under an arbitrary field redefinition 
which corresponds to a general coordinate transformation 
in the target manifold. In the perturbation theory, we assume that 
configurations of field variables are very close to a background field 
$\varphi_0$ corresponding to a vacuum, and expand the Lagrangian as a 
power series in $\delta\varphi^i=\varphi^i-\varphi^i_0$ to define the 
interaction Lagrangian. In general, the expansion coefficients of this 
power series are non-covariant quantities, 
like the Christoffel symbols $\Gamma^i{}_{jk}$. 
It is very convenient to choose a special 
coordinate system called {\it the Riemann normal coordinates}, 
in which all the expansion coefficients are covariant tensors. 
In this coordinate system, 
results of the perturbation theory are expressed in terms of covariant 
quantities, and reparameterization invariance becomes manifest. 
This is one reason why Riemann normal coordinates are widely 
used in the renormalization of sigma models~\cite{AFM}. 
The Riemann normal coordinates around $\varphi_0$ 
are usually defined as a 
coordinate system in which all geodesics passing through $\varphi_0$ 
are straight lines, 
and neighboring points are identified 
with tangent vectors at $\varphi_0$. 

Generally speaking, we have to solve the geodesic equation 
in order to find the coordinate transformation 
to the Riemann normal system. 
In this article, 
we propose a simple alternative algorithm to find the coordinate 
transformation to the normal coordinate system in the case that 
the nonlinear sigma models have $N=2$ supersymmetry in two dimensions. 
The existence of $N=2$ supersymmetry in two dimensions requires 
the target manifold to be a K\"{a}hler manifold~\cite{Zu}. 
Instead of using a metric, we rely heavily on 
the K\"{a}hler potential $K(\varphi,\varphi^*)$, 
which fixes the geometry 
of the K\"ahler manifold, to transform to Riemann normal coordinates.  
We call our normal coordinates 
``K\"{a}hler normal coordinates''.\footnote{ 
Although Clark and Love~\cite{CL} presented one such method, 
it requires an isometry of the target manifold. On the other hand, 
our method is valid for any K\"{a}hler manifold.
}
One of novel features of our method is that 
we do not need discussions of geodesics,  
in contrast to the real Riemann manifold~\cite{RNC}. 

Nonlinear sigma models with $N=2$ supersymmetry are very 
important tools to describe superstring theory in 
compactified space-time. Recently we found auxiliary 
field formulations for nonlinear sigma models on Hermitian 
symmetric spaces~\cite{HN}. We hope our new method together with 
the auxiliary field formulation will play a significant role 
in the non-perturbative analyses of these models. 

One can also apply the normal coordinate method 
to nonlinear sigma models in four dimensions. 
The $N=1$ supersymmetry in four dimensions has 
the same structure as the $N=2$ supersymmetry in two dimensions. 
(The latter is a direct dimensional 
reduction of the former.)  
$N=1$ supersymmetric nonlinear sigma models 
in four dimensions appear as 
low-energy effective theories describing 
(quasi-)Nambu-Goldstone bosons
when global symmetry is spontaneously broken 
with preserved supersymmetry~\cite{BKMU}. 
The low-energy theorems of two-body scattering amplitudes 
of these bosons are discussed in Ref.~\cite{HNOO}, 
where a K\"{a}hler normal coordinate expansion to fourth order 
is used. An expansion to higher orders is necessary for calculations of 
many-body scattering amplitudes. 

This paper is organized as follows. 
In section 2 we construct the K\"{a}hler normal coordinate expansion 
and present a theorem asserting that all the coefficients are covariant.
This method is applied to supersymmetric 
nonlinear sigma models in section 3.
We summarize the geometry of the K\"{a}hler manifold in Appendix A. 
A proof of the theorem is given in Appendix B.

%%%%%%%%%%%%%%%%%%%%
\section{K\"{a}hler normal coordinate} 
In the case of real Riemannian manifolds, we have to solve the 
geodesic equation to find the transformation to the 
Riemann normal coordinates.
We propose to use the K\"{a}hler potential to find the 
holomorphic transformation to the K\"{a}hler normal 
coordinates. If we expand the K\"{a}hler potential in 
a Taylor series about the origin, the coefficients are in general 
non-covariant quantities. After an appropriate coordinate 
transformation, all coefficients are expressed in terms of 
covariant quantities.
In Subsec.~\ref{4-th} we discuss the K\"{a}hler normal 
coordinates to fourth order. 
All non-covariant terms in the Taylor expansion of the K\"{a}hler 
potential are eliminated explicitly to this order.  
In Subsec.~\ref{all-order}, we generalize the coordinate 
transformation to all orders and give a theorem ensuring  
the covariance of all coefficients in the new coordinate system.  
A proof of the theorem is given in Appendix B, 
and we calculate several lower order coefficients 
to confirm their covariance in the remainder of 
Subsec.~\ref{all-order}. 

%%%%%%%%%%%%%%%
\subsection{The Fourth order}\label{4-th}
In this subsection, we discuss the fourth order 
K\"{a}hler normal coordinates. 
Let $\left( z^i,z^{*i} \right)$ be 
the general complex coordinate of 
a patch of a K\"{a}hler manifold. 
A K\"{a}hler manifold is characterized by  
a K\"{a}hler potential $K(z,z^*)$, which is 
defined in each coordinate patch of the manifold. 
Then the K\"{a}hler metric is given by 
\beq
 g_{ij^*}(z,z^*) = \del_i \del_{j^*} K(z,z^*), \label{Kahler-metric}
\eeq
where the differentiations are with respect to 
the coordinates $z^i$ and $z^{*j}$.
The metric is invariant under the K\"{a}hler transformation 
\beq
 K(z,z^*) \to K(z,z^*) + f(z) + f^*(z^*). \label{Kahler-tr.}
\eeq
Geometric quantities, such as the connection 
and the curvature, can be calculated from the metric 
and hence from the K\"{a}hler potential, 
as summarized in Appendix A. 

We frequently use convenient notation for partial derivatives. 
In this notation, indices preceded by a comma  
denote derivatives. For example, the definition of 
the K\"{a}hler metric (\ref{Kahler-metric}) is written 
$g_{ij^*}(z,z^*)=K,_{ij^*}$. 
Then the Taylor expansion of the K\"{a}hler potential 
around the coordinate origin $z^i=0$ is   
\beq
     K(z,z^*) 
&=& \sum_{N,M=0}^{\infty} 
    \1{N!M!} K,_{\, i_1 \cdots i_N j_1^* \cdots j_{M}^*}|_0 
    z^{i_1} \cdots z^{i_N} z^{*j_1} \cdots z^{*j_M} \non 
&=&  K|_0 + F(z) + F^*(z^*) \non
&& + g_{ij^*}|_0\, z^i z^{*j} 
   + \1{2}\Gamma_{i^*jk}|_0\, z^{*i} z^j z^k
   + \1{2} {\Gamma}_{ij^*k^*}|_0\, z^i z^{*j} z^{*k} \non
&& + \1{4} (R_{ij^*kl^*}
       + g_{mn^*}{{\Gamma}^m}_{ik}{{\Gamma}^{n^*}}_{j^*l^*})  
            |_0 \, z^i z^k z^{*j} z^{*l}\non
&& + \1{6}\del_k {\Gamma}_{l^*ij}|_0\, z^i z^j z^k z^{*l} 
   + \1{6}\del_{k^*} {\Gamma}_{li^*j^*}|_0\, 
          z^{*i} z^{*j} z^{*k} z^l + O(5) ,\label{TE of K}
\eeq 
where
\beq
 F(z) = \sum_{N=1}^{\infty} \1{N !} 
        K,_{\,i_1 \cdots i_N}|_0 z^{i_1} \cdots z^{i_N}
 = K,_{\,i}|_0 \,z^i 
 + \1{2} K,_{\,ij}|_0 \,z^i z^j + \cdots
\eeq
is holomorphic and can be eliminated 
by the K\"{a}hler transformation (\ref{Kahler-tr.}).
The expansion coefficients are identified with the connection 
${\Gamma^i}_{jk}$ or  the curvature tensor $R_{ij^*kl^*}$ by 
using the definition of these geometrical quantities summarized 
in Appendix A. The subscripts $|_0$ indicate that the values 
in question are evaluated at the origin $z^i=0$. 
(We sometimes omit ``0'' when the expansion point is obvious.)
With the exception of $g_{ij^*}$ and $R_{i^*jk^*l}$, 
all coefficients are non-covariant to this order. 
A holomorphic coordinate transformation to eliminate these 
non-covariant quantities is found without difficulty. 
Note that Eq.~(\ref{TE of K}) can be rewritten as 
\beq
 K(z,z^*) 
 &=& K| + F(z) + F^*(z^*) \non
 &+& g_{mn^*}(z^m + \1{2} {\Gamma^m}_{jk}|z^j z^k 
   + \1{6}g^{ml^*}\del_k \Gamma_{l^*ij}| z^i z^j z^k )\non 
 && \hs{5} \times (z^n + \1{2} {\Gamma^n}_{op}|z^o z^p 
   + \1{6}g^{nr^*} \del_q{\Gamma}_{r^*op}| z^o z^p z^q )^* \non
 &+& \1{4} R_{i^*jk^*l}| z^{*i} z^{*k} z^j z^l + O(5) .
\eeq
So by the {\it holomorphic} coordinate transformation 
\beq
  \omega^i = z^i + \1{2} {\Gamma^i}_{jk}|z^j z^k 
         + \1{6} g^{im^*} \del_l \Gamma_{m^*jk}| z^j z^k z^l  
  \label{hol.co.tr.}  ,
\eeq
it can be written as
\beq
 K(\omega,\omega^*) 
 = K| + \tilde{F}(\omega) + \tilde{F}^*(\omega^*) 
 + g_{ij^*}| \omega^i \omega^{*j} 
  + \1{4} R_{i^*jk^*l}| {\omega}^{*i} {\omega}^{*k} 
                         {\omega}^j {\omega}^l 
 + O(5) ,\label{KNC_4-th_order}
\eeq
where $\tilde F(\omega) \defeq F(z(\omega))$. 
This coordinate transformation is invertible to give 
$z^i = z^i (\omega) 
=\omega^i - \1{2} {\Gamma^i}_{jk}|\omega^j \omega^k
        + \cdots $.

Non-covariant quantities remain in $\tilde F(\omega)$.
Since the transformation (\ref{hol.co.tr.}) is 
{\it holomorphic}, $\tilde F(\omega)$ is still 
holomorphic and can be eliminated by a K\"{a}hler 
transformation (\ref{Kahler-tr.}).  
Therefore, all the expansion coefficients are expressed in terms of  
covariant quantities in the new coordinate system spanned by $\omega$. 
This coordinate system is the 
desired K\"{a}hler normal coordinates.
We can use normal coordinates 
defined about an arbitrary point ${z^i}_0$ by simply replacing 
$z^i$ by $z^i - {z^i}_0$ in the above expression.  
All coefficients are evaluated at ${z^i}_0$ 
in this case. 

It is useful to calculate some geometric quantities 
in these K\"{a}hler normal coordinates. 
The metric is calculated as  
\beq
  g_{ij^*} (\omega,\omega^*) 
 = K,_{\,ij^*} (\omega,\omega^*)
 = g_{ij^*}| 
+ R_{ij^*kl^*}| \omega^k \omega^{*l}+ O(3),
\eeq
and the inverse metric follows from 
$g^{ij^*}g_{kj^*} = \delta^i_k$ as 
\beq
 g^{ij^*}(\omega,\omega^*) = 
 g^{ij^*}|+ {R^{ij^*}}_{kl^*}| \omega^k \omega^{*l}+ O(3). 
\eeq
\ From Eq.~(\ref{KNC_4-th_order}), 
we find that the curvature tensor {\it at the origin} is simply   
\beq
 R_{ij^*kl^*}| = K,_{\,ij^*kl^*}|. 
\eeq
Because of the commutativity of the differentiation, we obtain 
nontrivial relations among components of 
the curvature tensor from this equation 
(see Eq.~(\ref{sym.of_R2})): 
\beq
 R_{ij^*kl^*} = R_{kj^*il^*} = R_{il^*kj^*}.
    \label{sym.of_R}
\eeq
Since these equations are covariant, 
they hold in any coordinate system.

%%%%%%%%%%%%%%%%%%%%
\subsection{K\"{a}hler normal 
coordinate to all orders}\label{all-order}
In this subsection 
we generalize the coordinate transformation 
(\ref{hol.co.tr.}) to all orders.  
We then give a theorem which states that 
coefficients in the new coordinates are covariant. 
(A proof is given in Appendix B.) 
We also explicitly express the sixth order coefficients 
in terms of the curvature and its covariant 
derivatives for definiteness.
The simple Taylor expansion is again
\beq
     K(z,z^*) 
&=& K| + F(z) + F^*(z^*) + g_{ij^*}| z^i z^{*j} \non 
&& + \sum_{N=3}^{\infty} \sum_{M=1}^{N-1} 
    \1{M!(N-M)!} K,_{\, i_1 \cdots i_M j_1^* \cdots j_{N-M}^*}| 
    z^{i_1} \cdots z^{i_M} z^{*j_1} \cdots z^{*j_{N-M}} \non  
&=& K| + F(z) + F^*(z^*) + g_{ij^*}|  z^i z^{*j} 
   + \1{2}\Gamma_{i^*jk}| z^{*i} z^j z^k
   + \1{2} {\Gamma}_{ij^*k^*}|  z^i z^{*j} z^{*k} \non
 &&+ \1{4} K,_{\,i^*jk^*l}| z^{*i} z^{*k} z^j z^l   
   + \1{6}\del_k {\Gamma}_{l^*ij}| z^i z^j z^k z^{*l} 
   + \1{6}\del_{k^*} {\Gamma}_{li^*j^*}|
          z^{*i} z^{*j} z^{*k} z^l  \non 
 &&+ \1{12} K,_{\,m ij^*kl^*}| z^m z^i z^k z^{*j} z^{*l}
   + \1{12} K,_{\,m^* ij^*kl^*}| z^i z^k z^{*m} z^{*j} z^{*l} \non
 &&+\cdots .\label{gen.co.exp.}
\eeq 
The expansion coefficients are expressed in terms of geometric 
quantities by using the formulas given in Appendix A.
As a generalization of Eq.~(\ref{hol.co.tr.}), 
we perform a coordinate transformation given by  
\beq
  \omega^i &=& 
    z^i + \sum_{N=2}^{\infty} 
    \1{N!} g^{ij^*} K,_{\,i_1\cdots i_N j^*}| 
           z^{i_1} \cdots z^{i_N} \non
 &=& z^i + \sum_{N=2}^{\infty} \1{N!} g^{ij^*} 
   \del_{i_3 \cdots i_N} \Gamma_{j^* i_1 i_2}| 
   z^{i_1} \cdots z^{i_N} \non
 &=& z^i + \1{2} {\Gamma^i}_{jk}|z^j z^k
  + \1{6}(\del_l {\Gamma^i}_{jk} 
       + {\Gamma^i}_{ml} {\Gamma^m}_{jk})| z^j z^k z^l + \cdots
  \label{co.tr.}
\eeq
in order to eliminate terms of the form 
$z^{i_1}z^{*j_1} \cdots z^{*j_N}$ or 
$z^{i_1} \cdots z^{i_N} z^{*j_1}\; (N\ge 2)$.
We thus obtain the expansion
\beq
     K(\omega,\omega^*) 
&=& K| + \tilde F(\omega) + \tilde F^*(\omega^*) 
   + g_{ij^*}| \omega^i \omega^{*j} \non 
&& + \sum_{N=4}^{\infty} \sum_{M=2}^{N-2} 
    \1{M!(N-M)!} K,_{\, i_1 \cdots i_M j_1^* \cdots j_{N-M}^*}| 
    \omega^{i_1} \cdots \omega^{i_M} 
    \omega^{*j_1} \cdots \omega^{*j_{N-M}} \non  
&=& K|+ \tilde F(\omega) + \tilde F^*(\omega^*) 
  + g_{ij^*}|  \omega^i \omega^{*j} 
  + \1{4} K,_{i^*jk^*l}|\omega^{*i}\omega^{*k} 
                        \omega^j \omega^l  \non 
 &&+ \1{12} K,_{m ij^*kl^*}| 
  \omega^m \omega^i \omega^k \omega^{*j} \omega^{*l}
   + \1{12} K,_{m^* ij^*kl^*}| 
  \omega^i \omega^k \omega^{*j} \omega^{*l} \omega^{*m}\non
 &&+\cdots , \label{KNC_exp.}
\eeq 
where all differentiations are with respect to 
the new coordinates, $\omega$. 
In the previous subsection, 
we found that the coefficients in 
the expansion~(\ref{KNC_4-th_order}) 
are covariant quantities. The following theorem is a 
generalization of this observation.

\medskip
{\bf Theorem.} 
All coefficients in the expansion (\ref{KNC_exp.})
are covariant.\medskip\\
We call such a coordinate system $\omega$ the   
``K\"{a}hler normal coordinates to all orders''.
We prove this theorem in Appendix B. 
In this subsection, as an illustration, 
we explicitly express the first 
several coefficients in terms of the curvature 
and the covariant derivatives.  
We refer to a tensor with $N$ holomorphic lower indices and 
$M$ anti-holomorphic lower indices as an $(N,M)$ tensor.
Since we have eliminated terms of the form
$\omega^{i_1}\cdots\omega^{i_N}\omega^{*j_1}$ by the holomorphic 
coordinate transformation (\ref{co.tr.}), the connection 
$\Gamma_{j_1^* i_1 i_2}$ differentiated any number of times 
with respect to the holomorphic coordinates $\omega$,  
\beq
  K,_{\,i_1\cdots i_N j_1^*} 
   = \del_{i_3} \cdots \del_{i_N} \Gamma_{j_1^* i_1 i_2}
   = \del_{i_3} \cdots \del_{i_N} (g_{kj_1^*} {\Gamma^k}_{i_1 i_2}), 
\eeq
vanishes at the origin. 
This implies that $g_{ij^*, i_1 \cdots i_N}|=0$, and thus we find 
\beq
 \del_{i_1} \cdots \del_{i_N}{\Gamma^i}_{jk}| = 0.  
\eeq
Hence if all of the covariant derivatives, acting on 
any tensor $T$, are holomorphic or anti-holomorphic, 
they become ordinary derivatives with respect to 
the coordinates at the origin: 
\beq
  D_{i_1}\cdots D_{i_N} T| = \del_{i_1}\cdots \del_{i_N} T|.  
\eeq
In particular, we have very simple formulas for 
the curvature tensor:   
\beq
   D_{i_1}\cdots D_{i_N} R_{ij^*kl^*}| 
   = K,_{\,i_1\cdots i_N ij^*kl^*}|  , \hs{5}
   D_{j_1^*}\cdots D_{j_N^*} R_{ij^*kl^*}| 
 = K,_{\,j_1^*\cdots j_N^* ij^*kl^*}| \label{d.cur.} .
\eeq
For example, the $(3,2)$ tensor 
\beq
 && \hs{5} D_m R_{ij^*kl^*} \non
 && = \del_m R_{ij^*kl^*} - {\Gamma^n}_{mi} R_{nj^*kl^*} 
                         - {\Gamma^n}_{mk} R_{ij^*nl^*} \non
 && = K,_{\,mij^*kl^*} 
 + g^{op^*} g^{qn^*} K,_{\,p^*qm} K,_{\,oj^*l} K,_{\,n^*ik} \non
 && - g^{on^*} (K,_{\,omj^*l^*} K,_{\,n^*ik} 
    + K,_{\,oj^*l^*} K,_{\,mn^*ik} 
    + K,_{\,n^*mi}R_{oj^*kl^*} 
    + K,_{\,n^*mk}R_{ij^*ol^*}) \hs{10}
 \label{(3,2)}  
\eeq
satisfies
\beq
 D_m R_{ij^*kl^*}| = K,_{\,mij^*kl^*}|.
\eeq
In the calculation of Eq.~(\ref{(3,2)}), 
we have used the formula
\beq
 \del_k g^{ij^*} = - g^{im^*} g^{lj^*} K,_{\,m^*lk}. 
  \label{del_g}
\eeq 
The symmetry property of the curvature, (\ref{sym.of_R}), 
derived in the previous subsection, can be generalized. 
When all of the covariant derivatives acting 
on the curvature are (anti-)holomorphic, 
all (anti-)holomorphic indices of the tensor 
are symmetric as a result of Eq.~(\ref{d.cur.}):   
\beq
 D_{i_1} \cdots D_{i_a} \cdots D_{i_N} R_{ij^*kl^*} 
 = D_{i_1} \cdots D_i \cdots D_{i_N} R_{i_aj^*kl^*}, 
 \hs{10} {\rm etc.}
\eeq
Again these equations are covariant, 
and hence they hold in any coordinate system.
This relation can also be shown by the relation  
$[D_i,D_j]=0$ and the Bianchi identity.\footnote{  
The Bianchi identity in a Riemann manifold is 
$D_{(m} R_{ij)kl} = 
D_m R_{ijkl} + D_j R_{mikl} + D_i R_{jmkl}=0$. 
In a K\"{a}hler manifold this becomes  
$D_m R_{ij^*kl^*} + D_{j^*} R_{mikl^*} + D_i R_{j^*mkl^*}
= D_m R_{ij^*kl^*} - D_i R_{mj^*kl^*}= 0$. 
Hence we obtain $D_m R_{ij^*kl^*} = D_i R_{mj^*kl^*}$. 
The commutativity, $[D_i,D_j]=0$, 
follows from Footnote \ref{fn}, below. 
}

We now examine whether other terms in the expansion (\ref{KNC_exp.}) 
are also covariant. For example, the lowest non-trivial coefficient 
is, $K,_{\,mn^*ij^*kl^*}$, of the $(3,3)$-type.  
To evaluate this term we calculate one of 
the $(3,3)$-tensors from Eq.~(\ref{(3,2)}):
\beq
 && \hs{5} D_{n^*}D_m R_{ij^*kl^*} \non
 &&= \underline{K,_{\,mn^*ij^*kl^*}} 
  - (g^{ot^*} g^{sp^*} g^{qr^*} + g^{op^*} g^{qt^*} g^{sr^*}) 
   K,_{\,st^*n^*} K,_{\,p^*qm} K,_{\,oj^*l} K,_{\,r^*ik}\non
 &&+ g^{ot^*}g^{sr^*} 
   [ K,_{\,n^*t^*sm} K,_{\,oj^*l} K,_{\,r^*ik}
   + K,_{\,t^*sm} K,_{\,n^*oj^*l} K,_{\,r^*ik}
   + K,_{\,t^*sm} K,_{\,oj^*l} K,_{\,n^*r^*ik}) \non
 && \hs{15} + K,_{\,st^*n^*} 
    (K,_{\,omj^*l^*} K,_{\,r^*ik} + K,_{\,oj^*l^*} K,_{\,mr^*ik}\non
 && \hs{32} 
    + K,_{\,r^*mi} R_{oj^*kl^*} 
    + K,_{\,r^*mk} R_{ij^*ol^*}) \;] \non 
 &&- g^{or^*} 
   ( K,_{\,n^*omj^*l^*}K,_{\,r^*ik} + K,_{\,n^*oj^*l^*}K,_{\,mr^*ik}
   + \underline{K,_{\,omj^*l^*} K,_{\,n^*r^*ik}} 
   + K,_{\,oj^*l^*} K,_{\,n^*mr^*ik} \non
 && \hs{8} + \underline{K,_{\,n^*r^*mi} R_{oj^*kl^*}} 
    + \underline{K,_{\,n^*r^*mk} R_{ij^*ol^*}} 
    + K,_{\,r^*mi} \del_{n^*} R_{oj^*kl^*} 
    + K,_{\,r^*mk} \del_{n^*} R_{ij^*ol^*}\non
 && \hs{8} + K,_{\,rn^*j^*} D_m R_{io^*kl^*} 
    + K,_{\,rn^*l^*} D_m R_{ij^*ko^*}). \label{D*DR}
\eeq
Here, only the underlined terms survive at the origin. 
We thus obtain a covariant expression of 
the coefficient $(3,3)$, 
\beq 
 K,_{\,mn^*ij^*kl^*}| = D_{n^*} D_m R_{ij^*kl^*}| 
 + g^{or^*} R_{o(j^*ml^*} R_{in^*k)_{\rm h} r^*}| ,\label{(3,3)}
\eeq
where $(\cdots)_{\rm h}$ denotes cyclic permutation 
with respect to the holomorphic indices.
(For example, 
$A_{(ij^*kl)_{\rm h}} = A_{ij^*kl} + A_{lj^*ik} + A_{kj^*li}$.)
Note that this expression is not unique. 
For example, it can also be expressed as  
\beq
 K,_{\,mn^*ij^*kl^*}| = D_m D_{n^*} R_{ij^*kl^*}| 
 + g^{or^*} R_{o(j^*ml^*} R_{in^*k)_{\rm ah}r^*}| ,
  \label{(3,3)-2}
\eeq
where $(\cdots)_{\rm ah}$ denotes cyclic permutation 
with respect to the anti-holomorphic indices.~\footnote{  
These two expressions, Eqs.~(\ref{(3,3)}) and (\ref{(3,3)-2}),  
are related by a formula valid for any tensor $T$, 
\beq
 [D_A,D_B] \,T_{C_1 \cdots C_n} 
= \sum_{a=1}^n 
  {R_{ABC_a}}^D T_{C_1 \cdots C_{a-1} D C_{a+1} \cdots C_n},
   \label{com.D-D}
\eeq  
where Roman uppercase letters are used for both 
the holomorphic and anti-holomorphic indices. 
Note that $[D_i,D_j] = [D_{i^*},D_{j^*}] =0$ 
as a result of the K\"{a}hler property.  
Hence we can define a ``normal ordering'' 
by putting $D$ to the right of $D_*$ to obtain 
the unique expressions. We use this expression in our proof 
of the theorem.\label{fn}
}
The right-hand sides of Eqs.~(\ref{(3,3)}) and (\ref{(3,3)-2})
are manifestly symmetric on either the anti-holomorphic 
or holomorphic indices, but not both.   
The expressions symmetric with respect to 
both the holomorphic and anti-holomorphic 
indices are
\beq 
  K,_{\,mn^*ij^*kl^*}| 
 &=& \1{3} [D_{(n^*} D_m R_{ij^*kl^*)_{\rm ah}} 
 + g^{or^*} R_{o(j^*ml^*} R_{in^*k)r^*}]| \non 
 &=& \1{3} [D_{(m} D_{n^*} R_{ij^*kl^*)_{\rm h}}  
 + g^{or^*} R_{o(j^*ml^*} R_{in^*k)r^*}]|, \label{(3,3)-3}
\eeq 
where $(\cdots)$ denotes cyclic permutation 
with respect to both the holomorphic and anti-holomorphic indices, 
applied independently.\footnote{From this equation, 
we obtain the nontrivial identity 
$D_{(n^*} D_m R_{ij^*kl^*)_{\rm ah}} 
= D_{(m} D_{n^*} R_{ij^*kl^*)_{\rm h}}$. 
This can be also proved by the formula (\ref{com.D-D}). 
}  

In summary, from Eqs.~(\ref{d.cur.}) and (\ref{(3,3)-3}),  
the manifestly covariant expression of 
the K\"{a}hler normal coordinate expansion 
to sixth order can be written as 
\beq
 && \hs{5}   K(\omega,\omega^*) \non
 && =  K| + \tilde F(\omega) + \tilde F^*(\omega^*) 
   + g_{ij^*}| \omega^i \omega^{*j}  
   + \1{4} R_{ij^*kl^*}|\omega^i\omega^k 
                        \omega^{*j} \omega^{*l} \non 
 &&+ \1{12} D_m R_{ij^*kl^*}| 
    \omega^m \omega^i \omega^k \omega^{*j} \omega^{*l}
   + \1{12} D_{m^*}R_{ij^*kl^*}| 
    \omega^i \omega^k \omega^{*j} \omega^{*l} \omega^{*m}\non
 && + \1{24} D_n D_m R_{ij^*kl^*}| 
    \omega^n\omega^m\omega^i\omega^k\omega^{*j}\omega^{*l}
   + \1{24} D_{n^*}D_{m^*}R_{ij^*kl^*}| 
    \omega^i\omega^k\omega^{*j}\omega^{*l}\omega^{*m}\omega^{*n}\non
 && + \1{108}(D_{(n^*} D_m R_{ij^*kl^*)_{\rm ah}} 
         + g^{or^*} R_{o(j^*ml^*} R_{in^*k)r^*})|
    \omega^m\omega^i\omega^k\omega^{*j}\omega^{*l}\omega^{*n} 
   + O(7). \hs{8}\label{KNC_exp-6th}
\eeq 
By the same procedure, in principle, one can obtain 
covariant expressions of the expansion to any desired order.
All the coefficients are guaranteed to be covariant 
by the theorem.
 
In the rest of this section,  
we give K\"{a}hler normal coordinate expansions of 
some geometric quantities. 
The general expression of the K\"{a}hler metric in 
the K\"{a}hler normal coordinates to all orders is
\beq
  g_{ij^*} (\omega,\omega^*) 
  = g_{ij^*}|  
   + \sum_{N=1}^{\infty} \sum_{M=1}^{\infty} 
    \1{N!M!} K,_{\, ij^* i_1 \cdots i_N 
                      j_1^* \cdots j_M^*}| 
    \omega^{i_1} \cdots \omega^{i_N} 
    \omega^{*j_1} \cdots \omega^{*j_M}. 
\eeq
Note that $g_{ij^*},_{\,i_1 \cdots i_n}| 
= g_{ij^*},_{\,j_1^* \cdots j_n^*}|= 0$. 
The manifestly covariant expression of 
the expansion of the metric to fourth order is
\beq
 &&\hs{5} g_{ij^*}(\omega,\omega^*) \non
 &&= g_{ij^*}|   
   + R_{ij^*kl^*}|\omega^k \omega^{*l} 
  + \1{2} D_m R_{ij^*kl^*}| 
    \omega^m \omega^k \omega^{*l}
   + \1{2} D_{m^*}R_{ij^*kl^*}| 
    \omega^k \omega^{*l} \omega^{*m}\non
 && + \1{6} D_n D_m R_{ij^*kl^*}| 
    \omega^n\omega^m\omega^k\omega^{*l} 
    + \1{6} D_{n^*}D_{m^*}R_{ij^*kl^*}| 
    \omega^k\omega^{*l}\omega^{*m}\omega^{*n}\non
 && + \1{12}(D_{(n^*} D_m R_{ij^*kl^*)_{\rm ah}} 
         + g^{or^*} R_{o(j^*ml^*} R_{in^*k)r^*})|
    \omega^m\omega^k\omega^{*l}\omega^{*n} 
    + O(5) \label{metric_exp-4th}.
\eeq 
The inverse metric in the normal coordinate expansion 
can be calculated order by order from 
the definition $g^{ij^*}g_{j^*k} = \delta^i_k$. 
The expansion to fourth order is
\beq
 && \hs{5} g^{ij^*}(\omega,\omega^*) \non
 && = g^{ij^*}|   
   + {R^{ij^*}}_{kl^*}|\omega^k \omega^{*l} 
   + \1{2} D_m {R^{ij^*}}_{kl^*}| 
    \omega^m \omega^k \omega^{*l}
   + \1{2} D_{m^*} {R^{ij^*}}_{kl^*}| 
    \omega^k \omega^{*l} \omega^{*m}\non
 && + \1{12} D_n D_m {R^{ij^*}}_{kl^*}| 
    \omega^n\omega^m\omega^k\omega^{*l} 
    + \1{6} D_{n^*}D_{m^*}{R^{ij^*}}_{kl^*}| 
    \omega^k\omega^{*l}\omega^{*m}\omega^{*n} \non
 &&   - \1{4} g^{iq^*} g^{j^*p}(D_{(n^*} D_m R_{pq^*kl^*)_{\rm ah}} 
         + g^{or^*} R_{o(q^*ml^*} R_{pn^*k)r^*} 
  - g^{or^*} R_{oq^*(mn^*}R_{kl^*)pr^*}) | \non
  && \hs{20}\times  \omega^m\omega^k\omega^{*l}\omega^{*n} 
   + O(5) \label{inverse-metirc_exp}.
\eeq 
The expansion of the connection can be calculated as   
\beq
 && \hs{5} \Gamma_{j^*ik}(\omega,\omega^*) 
   = K,_{\,ikj^*}(\omega,\omega^*) 
   =  g_{ij^*},_k(\omega,\omega^*)  \non
 &&= R_{ij^*kl^*}| \omega^{*l} 
   + D_m R_{ij^*kl^*}| 
    \omega^m \omega^{*l}
   + \1{2} D_{m^*}R_{ij^*kl^*}| 
     \omega^{*l} \omega^{*m}\non
 && + \1{2} D_n D_m R_{ij^*kl^*}| 
     \omega^n\omega^m\omega^{*l} 
    + \1{6} D_{n^*}D_{m^*}R_{ij^*kl^*}| 
     \omega^{*l}\omega^{*m}\omega^{*n}\non
 && + \1{6}(D_{(n^*} D_m R_{ij^*kl^*)_{\rm ah}} 
         + g^{or^*} R_{o(j^*ml^*} R_{in^*k)r^*})|
    \omega^m\omega^{*l}\omega^{*n}  
   + O(4) \label{connection_exp}.
\eeq 
Note that each term has at least one 
anti-holomorphic factor, $\omega^*$, 
and hence the holomorphic derivatives of the connection
are zero at the origin.  
The curvature tensor can be calculated to 
second order from Eqs.~(\ref{inverse-metirc_exp}) 
and (\ref{connection_exp}):
\beq
  && \hs{5} R_{ij^*kl^*}(\omega,\omega^*) \non
 && = R_{ij^*kl^*}|
   + D_m R_{ij^*kl^*}| \omega^m 
   + D_{m^*}R_{ij^*kl^*}| \omega^{*m}\non
 && + \1{2} D_n D_m R_{ij^*kl^*}| \omega^n\omega^m 
    + \1{2} D_{n^*}D_{m^*}R_{ij^*kl^*}| 
      \omega^{*m}\omega^{*n}\non
 && + \1{3}(D_{(n^*} D_m R_{ij^*kl^*)_{\rm ah}} 
       + g^{or^*} R_{o(j^*ml^*} R_{in^*k)r^*}
       - g^{or^*} R_{oj^*ml^*} R_{in^*kr^*})|
      \omega^m \omega^{*n}\non 
 &&  + O(3). \label{curvature_exp}
\eeq 

%%%%%%%%%%%%
\section{Applications to supersymmetric nonlinear sigma models}
$N=1$ ($N=2$) supersymmetry in four (two) dimensions 
requires the target manifold of nonlinear sigma models 
to be a K\"{a}hler manifold~\cite{Zu}. 
We first present the derivation appearing in Ref.~\cite{WB} 
of the Lagrangian of supersymmetric nonlinear sigma models 
in the general coordinates. 
After a brief remark on a field redefinition,  
we apply K\"{a}hler normal coordinates  
to background field methods. 

%%%%%%%%%%%%%%
\subsection{Review of the chiral model}
Chiral superfields satisfying 
the constraint $\bar D_{\dot{\alpha}} \Phi = 0$ 
are given by  
\beq
 &&\Phi^i(x, \th,\thb) = 
 \Phi^i(y, \th) = \ph^i(y) 
  + \sqrt 2 \th \psi^i(y) + \th\th F^i(y),\\ 
 &&y^{\mu} = x^{\mu} + i \th \sig^{\mu} \thb , \hs{10}
 \bar D_{\dot{\alpha}} = 
 - {\del \over \del \bar \th^{\dot{\alpha}}}.
\eeq
The general D-term Lagrangian of the chiral superfields 
can be written as 
\beq
 {\cal L} = \int d^4 \th K(\Phi,\Phi\dagg) ,
\eeq
where the K\"{a}hler potential $K$ is a real function.  
To calculate the Lagrangian written 
in terms of component fields, 
we expand the K\"{a}hler potential as in Eq.~(\ref{gen.co.exp.}):  
\beq
 K = \sum_{N,M=0}^{\infty} \1{N! M!} 
   K,_{\,i_1\cdots i_N j_1^* \cdots  j_M^*}|_0 
  \Ph^{i_1}\Ph^{i_2} \cdots \Ph^{i_N}
   \Ph^{\dagger j_1}\Ph^{\dagger j_2} \cdots \Ph^{\dagger j_M}.
  \label{Tayler-exp.}
\eeq
We define 
\beq
 K_{NM} = \Ph^{i_1}\Ph^{i_2} \cdots \Ph^{i_N}
     \Ph^{\dagger j_1}\Ph^{\dagger j_2} 
     \cdots \Ph^{\dagger j_M}. \label{poly.}
\eeq
Its D-term can be calculated as 
\beq
 [K_{NM}]_{\rm D}  
 &=& {\del^2 K_{NM}(\ph,\ph^*) 
       \over \del\ph^i \del\phs{j}} F^i F^{*j}
     - \1{2} {\del^3 K_{NM}(\ph,\ph^*) 
              \over \del\ph^i \del\phs{j} \del\phs{k}}
     F^i \psb^j \psb^k \non
 &&  - \1{2} {\del^3 K_{NM}(\ph,\ph^*) 
               \over \del\phs{i} \del\ph^j \del\ph^k}
     F^{*i} \ps^j \ps^k 
    + \1{4} {\del^4 K_{NM}(\ph,\ph^*) \over
 \del\ph^i \del\ph^j \del\phs{k} \del\phs{l}}
    \ps^i \ps^j \psb^k \psb^l \non
 && + {\del^2 K_{NM}(\ph,\ph^*) \over \del\ph^i \del\phs{j}} 
    \del_{\mu}\ph^i \del^{\mu}\phs{j}
    + i{\del^2 K_{NM}(\ph,\ph^*) \over \del\ph^i \del\phs{j}}
    \psb^j\sigb^{\mu}\del_{\mu}\ps^i \non
 && +i {\del^3 K_{NM}(\ph,\ph^*) 
            \over \del\ph^i \del\ph^j \del\phs{k}}
    (\psb^k \sigb^{\mu} \ps^i) \del_{\mu}\ph^j . \label{formula}
\eeq
Here we have used the equations 
\beq
 &&K_{N0} (\Phi) 
 = K_{N0} (\ph) + \sqrt{2} \th \psi^i 
    {\partial K_{N0}(\ph) \over \partial\ph^i}\non
 && \hs{20}
  + \th\th \left(
    F^i {\partial K_{N0}(\ph) \over \partial\ph^i} 
     - \1{2}\ps^i \ps^j 
   {\partial^2 K_{N0} (\ph) \over \partial\ph^i \partial\ph^j}
   \right),  \label{chiral-product}\\
  &&[\Phi^i \Phi^{j\dagger}]_{\rm D} 
  = F^i F^{*j} - \1{4} \ph^i \Box \ph^{*j} 
  - \1{4} \Box \ph^i \ph^{*j} 
  + \1{2} \del_{\mu}\ph^i \del^{\mu}\ph^{*j} \non
 && \hs{20}
  - \2{2} \del_{\mu} \psb^j \sigb^{\mu}\ps^i
  + \2{2} \psb^j \sigb^{\mu} \del_{\mu}\ps^i , 
\eeq
and partial integration.\footnote{
This partial integration can be carried out,   
since the coefficients of Eq.~(\ref{Tayler-exp.}) 
are constant.
}
\ From Eqs.~(\ref{Tayler-exp.}) and (\ref{formula}), 
the general Lagrangian of chiral superfields 
can be written as 
\beq
 &&{\cal L} 
 = g_{ij^*}F^iF^{*j}
    - \1{2} g_{im^*} \Gamma^{m^*}_{j^*k^*}F^i \psb^j \psb^k 
    - \1{2} g_{mi^*} \Gamma^m_{jk}F^{*i} \ps^j \ps^k \non
 &&\hs{5}
    + g_{ij^*}\del_{\mu}\ph^i \del^{\mu}\phs{j} 
    + ig_{ij^*}\psb^j\sigb^{\mu} \del_{\mu}\ps^i 
    + ig_{lk^*} \Gamma^l_{ij} \psb^k \sigb^{\mu}\ps^i
                              \del_{\mu}\ph^j \non
 &&\hs{5}
    + \1{4} g_{ij^*,kl^*} \ps^i\ps^k\psb^j\psb^l.
   \label{general_kahler}
\eeq
The equation of motion of $F^i$ reads  
\beq
 F^i = \1{2} {\Gamma^i}_{jk}(\ph,\ph^*) \psi^j \psi^k. 
\eeq
By substituting this back into Eq.~(\ref{general_kahler}), 
we obtain the Lagrangian of 
the supersymmetric nonlinear sigma model 
in the component fields,   
\beq 
{\cal L} = g_{ij^*}(\ph,\ph^*)\del_{\mu}\ph^i \del^{\mu}\phs{j}
 + ig_{ij^*}(\ph,\ph^*)\psb^j \sigb^{\mu} (D_{\mu} \ps)^i 
  + \1{4} R_{ij^*kl^*}(\ph,\ph^*) \ps^i\ps^k \psb^j\psb^l,\non
  \label{snlsm}
\eeq
where $D_{\mu}$ on the fermion is 
a pull-back of the covariant derivative 
on target manifolds, 
where the fermion behaves like a tangent vector 
(see Eq.~(\ref{component-field_redef.}), below):
\beq 
 (D_{\mu} \ps)^i = \del_{\mu} \psi^i 
 + \del_{\mu}\ph^j {\Gamma^i}_{jk}(\ph,\ph^*) \psi^k. 
  \label{cov.deriv._fermion} 
\eeq

%%%%%%%%%%%
\subsection{Field redefinition of chiral superfields}
Before proceeding to discussion of 
the K\"{a}hler normal coordinates of 
nonlinear sigma models, 
we discuss a field redefinition of chiral superfields 
as a general coordinate transformation 
on target manifolds.
Since a holomorphic function of chiral superfields
$\Phi^i (x,\th,\thb)$ ($i=1,\cdots,n$) is a chiral superfield, 
new fields $\Phi^{\prime \,i}(x,\th,\thb)$ defined by
\beq
  \Phi^{\prime \,i}(x,\th,\thb) 
  = f^i (\Phi^j (x,\th,\thb)) \label{chiral-field_redef.}
\eeq 
are chiral superfields and can be used as coordinates of 
the K\"ahler manifold.
The right-hand side can be written 
in component fields from Eq.~(\ref{chiral-product}) as 
\beq
 && f^i (\Phi (y,\th)) 
  = f^i(\ph(y)) 
   + \sqrt 2 \th \psi^j {\del f^i (\ph)\over \del \ph^j}(y) \non
 && \hs{30} 
   + \th\th \left(F^j {\del f^i (\ph)\over \del \ph^j}(y) 
      - \1{2} \psi^j\psi^k  
        {\del^2 f^i (\ph)\over \del \ph^j \del \ph^k}(y) 
       \right).
\eeq
The field redefinitions of the component fields are
\beq
  && \ph^{\prime \, i}(x) = f^i(\ph(x)),  \non
  && \psi^{\prime \, i}(x) = 
   {\del f^i (\ph(x))\over \del \ph^k} \psi^j(x), \non
  && F^{\prime \, i}(x) = 
     {\del f^i (\ph(x))\over \del \ph^j} F^j(x) 
    - \1{2}{\del^2 f^i (\ph(x))\over \del \ph^j \del \ph^k} 
      \psi^j\psi^k(x) . \label{component-field_redef.} 
\eeq
Note that the field dependences on $x$ are 
the same as those on $y$,  
since the relation $y = x + i \th \sig \thb$ 
includes $\th$ and $\thb$.  
The first equation represents 
a general coordinate transformation, 
whereas the second equation implies  
that the fermions transform as a tangent vector 
on the target manifold, 
as expressed by Eq.~(\ref{cov.deriv._fermion}). 

For later use, 
we point out the field definition 
(\ref{chiral-field_redef.}) can be generalized to
\beq
 \Phi^{\prime \,i}(y,\th) 
  = f^i (\Phi^j (y,\th), \ph_0(y)) ,
    \label{gen.chiral-field_redef.}
\eeq
where $\ph_0(y)$ is an additional bosonic field. 
We consider $\ph_0$ as a background field 
in the next subsection. 
Note that the bosonic field $\ph_0$ can 
depend on $y$ but not on $x$. 
This is because we can preserve chirality: 
$\bar D_{\dot{\alpha}} \Phi^i=0$ implies 
$\bar D_{\dot{\alpha}} \Phi^{\prime\, j}=0$, 
since the spinor derivative 
$\bar D_{\dot{\alpha}} = 
 - {\del \over \del \bar \th^{\dot{\alpha}}}$ 
does not include $y$ 
in the $y$-representation. 
Transformations of the component fields are given simply by
\beq
  && \ph^{\prime \, i}(x) = f^i(\ph(x),\ph_0(x)),  \non
  && \psi^{\prime \, i}(x) = 
   {\del f^i (\ph(x),\ph_0(x))\over \del \ph^k} \psi^j(x), \non
  && F^{\prime \, i}(x) = 
     {\del f^i (\ph(x),\ph_0(x))\over \del \ph^j} F^j(x) 
  - \1{2}{\del^2 f^i (\ph(x),\ph_0(x)) 
          \over \del \ph^j \del \ph^k} 
      \psi^j\psi^k(x) . \label{gen.comp.-field_redef.} 
\eeq
The bosonic fields depending on 
$y$ and $x$ are related as 
\beq 
  \ph_0(y) = \ph_0(x) + \del_{\mu} \ph_0 (i \th\sig^{\mu}\thb) 
  + \1{4} \Box \ph_0(x) \th\th\thb\thb,  
  \label{ph_in_y-x}
\eeq 
and the difference between $\ph_0(y)$ and $\ph_0(x)$ 
contains at least a term proportional to $\th$ and $\thb$. 
The transformation (\ref{gen.chiral-field_redef.}) 
may depend on a bosonic field through 
an arbitrary tensor (or non-tensor) 
$T_{i_1 \cdots j_1^* \cdots}(\ph_0(y),{\ph_0}^*(y))$ 
on the target manifold.  
It can be expanded around $\ph_0(x)$ as
\beq 
 && T(\ph_0(y),{\ph_0}^*(y)) \non
 &=& T\left( 
     \ph_0(x) + \del_{\mu} \ph_0 (i \th\sig^{\mu}\thb) 
     + \1{4} \Box \ph_0(x) \th\th\thb\thb , \mbox{conj.} 
    \right) \non 
 &=& T|_{\ph_0(x)} 
    + (\del_{\mu}{\ph_0}^i(x) 
     \del_i T|_{\ph_0(x)}  
    - \del_{\mu}{\ph_0}^{*i}(x) 
     \del_{i^*} T|_{\ph_0(x)})
     (i\th\sig^{\mu}\thb) \non
  &+& \left( \1{4} \del_{\mu}{\ph_0}^i(x) \del^{\mu}{\ph_0}^j(x) 
     \del_i\del_j T|_{\ph_0(x)} 
      + \1{4} \del_{\mu}{\ph_0}^{*i}(x) 
              \del^{\mu}{\ph_0}^{*j}(x) 
     \del_{i^*}\del_{j^*} T|_{\ph_0(x)} \right.\non
  &&\left. 
    - \1{2} \del_{\mu}{\ph_0}^i(x) \del^{\mu}{\ph_0}^{*j}(x) 
     \del_i\del_{j^*} T|_{\ph_0(x)} \right) \th\th\thb\thb, 
  \label{y-x}
\eeq 
and the difference between $T|_{\phi_0(y)}$ and $T|_{\phi_0(x)}$ 
contains at least a term proportional to $\th$ and $\thb$.
In the next subsection, 
$\ph_0(x)$ is regarded as a background field. 

%%%%%%%%%%%%%%%%%%
\subsection{Nonlinear sigma model 
in the K\"{a}hler normal coordinate}
We now apply the results of the previous section to 
background field methods in sigma models. 
The dynamics are described by quantum fluctuations  
around a vacuum expectation value, given by
\beq
 \left<\Ph(y,\th) \right> 
  = \left<\ph(y) \right> = \ph_0(y).
\eeq
Here we consider a bosonic background 
and assume that $\left<F \right> =0$,  
so that supersymmetry is unbroken. 
Note that the background depends on $y$ but not $x$,  
as clarified below. 
The relation with the bosonic background in 
the ordinary coordinates $x$ is Eq.~(\ref{ph_in_y-x}).
We replace the complex coordinates $z^i - {z_0}^i$ 
of the K\"{a}hler manifolds in the last section with 
chiral superfields 
\beq
 \Delta \Phi^i(y,\th) \defeq 
  \Phi^i(y,\th) - {\ph_0}^i(y).
\eeq  
As a generalization of Eq.~(\ref{co.tr.}) to superfields, 
we perform the coordinate transformation to 
the K\"{a}hler normal coordinates $\xi^i (x,\th,\thb)$:  
\beq
 && \xi^i (y,\th) 
  = \xi^i (\Delta \Phi^j (y,\th), \ph_0(y) )   \non
 &=& \Delta \Phi^i + \sum_{N=2}^{\infty} 
    \1{N!} g^{ij^*} K,_{\,i_1\cdots i_N j^*}|_{\ph_0(y)} 
    \Delta \Phi^{i_1} \cdots \Delta \Phi^{i_N} \non
 &=& \Delta \Phi^i + \1{2} {\Gamma^i}_{jk}|_{\ph_0(y)}
     \Delta \Phi^j \Delta \Phi^k
  + \1{6}g^{im^*}\del_l {\Gamma}_{m^*jk}|_{\ph_0(y)}
 \Delta \Phi^j \Delta \Phi^k \Delta \Phi^l + \cdots. \hs{5}
 \label{field_redef.}
\eeq
Note that the two sets of chiral superfields 
$\xi^i(x,\th,\thb)$ and $\Delta \Phi^i(x,\th,\thb)$ 
have the same chirality: 
$\bar D_{\dot{\alpha}} \Delta \Phi^i=0$ implies 
$\bar D_{\dot{\alpha}} \xi=0$, as discussed 
in the previous subsection. 
This is because 
the coefficients are evaluated with  
the {\it bosonic} background 
$\Phi(y,\th) = \ph_0(y)$.\footnote{ 
If we consider background {\it superfields} $\Phi_0$,
two sets of superfields cannot be simultaneously chiral,  
since $K,_{\,i_1\cdots i_N j_1^*}(\Phi_0, {\Phi_0}\dagg)$ 
possesses chiral and anti-chiral superfields~\cite{Sp}.  
}
The bosonic and fermionic parts of Eq.~(\ref{field_redef.}) 
in $x$ are obtained from 
Eq.~(\ref{gen.comp.-field_redef.}) to give 
\beq 
 && \ph^i_{\xi}(x) =  
   \ph^i_{\Delta\Phi}(x) + \sum_{N=2}^{\infty} 
    \1{N!} g^{ij^*} K,_{\,i_1\cdots i_N j^*}|_{\ph_0(x)} 
    \ph^{i_1}_{\Delta\Phi}(x)\cdots 
       \ph^{i_N}_{\Delta\Phi}(x), \label{boson_redef.}\\
 && \psi^i_{\xi}(x) =
   \psi^i_{\Delta\Phi}(x) + 
   \sum_{N=1}^{\infty} 
    \1{N!} g^{ij^*} K,_{\,i_1\cdots i_N k j^*}|_{\ph_0(x)} 
    \ph^{i_1}_{\Delta\Phi}(x)\cdots 
       \ph^{i_N}_{\Delta\Phi}(x) 
    \psi^k_{\Delta\Phi}(x) , \hs{10} 
\eeq
respectively. 
Here we have set 
$\Delta \Phi(y,\th) = 
\ph_{\Delta \Phi}(y) + \sqrt 2 \th \psi_{\Delta \Phi}(y)
+ \th\th F_{\Delta \Phi}(y)$ and 
$\xi(y,\th) = \ph_{\xi}(y) + \sqrt 2 \th \psi_{\xi}(y)
+ \th\th F_{\xi}(y)$.

The same expansion as in Eq.~(\ref{KNC_exp.}) is obtained 
by the transformation (\ref{field_redef.}). 
We thus obtain a K\"{a}hler normal coordinate expansion 
of the Lagrangian of supersymmetric nonlinear sigma models, 
which is manifestly invariant 
under the supersymmetry transformation and 
the general coordinate transformation:    
\beq
 {\cal L} = \int d^4\th 
    \sum_{N=4}^{\infty} \sum_{M=2}^{N-2} 
    \1{M!(N-M)!} 
    K,_{\, i_1 \cdots i_M j_1^* \cdots j_{N-M}^*}|_{\ph_0(y)} 
    \xi^{i_1} \cdots \xi^{i_M} 
    \xi^{\dagger j_1} \cdots \xi^{\dagger j_{N-M}}, 
    \label{Lag._in_KNC}
 \non 
\eeq
where the covariance of the coefficients is ensured by 
the theorem in the previous section. 

The Lagrangian in terms of the component fields 
can be calculated in the same way as Eq.~(\ref{snlsm}), 
by noting that coefficients are not constant 
and we must calculate a product 
of Eq.~(\ref{y-x}) with 
$T = K,_{\,i_1 \cdots j_1^* \cdots}|_{\ph_0(y)}$ 
and Eq.~(\ref{poly.}) 
before integration over $\th$. 
Instead, we can integrate over $\th$ first, 
and then transform to the K\"{a}hler normal coordinates 
at the level of the component fields.\footnote{
This is ensured by the fact that 
field redefinitions of chiral superfields 
reduce to field redefinitions of 
components fields of bosons and fermions,  
as seen in Eq.~(\ref{gen.comp.-field_redef.}).
} 
To do this, we must calculate
\beq
 \del_{\mu}\ph^i 
 = \del_{\mu}(\ph_0^i + \Delta \ph^i) 
 = \del_{\mu}\ph_0^i + 
   \del_{\mu}\left(\ph_{\xi}^i  -\1{2} {\Gamma^i}_{jk}|_{\ph_0} 
     \ph_{\xi}^j \ph_{\xi}^k + \cdots\right), 
  \label{non-const.BG}
\eeq
where the inverse transformation of 
Eq.~(\ref{boson_redef.}) is needed. 

In the case of a constant background, $\del \ph_0 = 0$, 
the integration over $\th$ in (\ref{Lag._in_KNC}) 
can be performed easily. 
At the component level, 
we do not need Eq.~(\ref{non-const.BG}).  
An expansion to sixth order 
can be obtained by substituting 
Eqs.~(\ref{metric_exp-4th}), (\ref{connection_exp}) 
and (\ref{curvature_exp}) into Eq.~(\ref{snlsm}) as  
(we omit the subscript $\xi$)
\beq
 &&{\cal L}
   = g_{ij^*}(\ph,\ph^*) \del_{\mu}\ph^i \del^{\mu}\phs{j}
   + i g_{ij^*}(\ph,\ph^*)\psb^j\sigb^{\mu}\del_{\mu}\ps^i
   + i \Gamma_{j^*ik}(\ph,\ph^*) 
   \del_{\mu}\ph^k \psb^j\sigb^{\mu}\ps^i \non
 &&\hs{5} 
   + \1{4} R_{ij^*kl^*}(\ph,\ph^*) \ps^i\ps^k \psb^j\psb^l,\non  
 && = g_{ij^*}|\del_{\mu}\ph^i \del^{\mu}\ph^{*j}
    +ig_{ij^*}|\psb^j \sigb^{\mu} \del_{\mu}\ps^i \non 
 &&+ \left[ 
      R_{ij^*kl^*}|\ph^k \ph^{*l} 
   + \1{2} D_m R_{ij^*kl^*}| 
    \ph^m \ph^k \ph^{*l}
   + \1{2} D_{m^*}R_{ij^*kl^*}| 
    \ph^k \ph^{*l} \ph^{*m} \right. \non
 && \left.\hs{5} 
    + \1{6} D_n D_m R_{ij^*kl^*}| 
    \ph^n\ph^m\ph^k\ph^{*l} 
    + \1{6} D_{n^*}D_{m^*}R_{ij^*kl^*}| 
    \ph^k\ph^{*l}\ph^{*m}\ph^{*n}\right. \non
 && \left.\hs{5} 
    + \1{12}(D_{(n^*} D_m R_{ij^*kl^*)_{\rm ah}} 
         + g^{or^*} R_{o(j^*ml^*} R_{in^*k)r^*})|
    \ph^m\ph^k\ph^{*l}\ph^{*n} \right] \non
 && \hs{10} \times 
    (\del_{\mu}\ph^i \del^{\mu}\ph^{*j}
         + i \psb^j \sigb^{\mu} \del_{\mu}\ps^i) \non
 && + i \left[ R_{ij^*kl^*}| \ph^{*l} 
   + D_m R_{ij^*kl^*}| 
    \ph^m \ph^{*l}
   + \1{2} D_{m^*}R_{ij^*kl^*}| 
     \ph^{*l} \ph^{*m} \right.\non
 && \left.\hs{5} 
    + \1{2} D_n D_m R_{ij^*kl^*}| \ph^n\ph^m\ph^{*l}  
    + \1{6} D_{n^*}D_{m^*}R_{ij^*kl^*}| 
            \ph^{*l}\ph^{*m}\ph^{*n} \right. \non
 && \left.\hs{5} 
    + \1{6}(D_{(n^*} D_m R_{ij^*kl^*)_{\rm ah}} 
         + g^{or^*} R_{o(j^*ml^*} R_{in^*k)r^*})|
    \ph^m\ph^{*l}\ph^{*n}\right]   
   \del_{\mu}\ph^k \psb^j\sigb^{\mu}\ps^i \non
 && + \1{4} \left[  R_{ij^*kl^*}|
   + D_m R_{ij^*kl^*}| \ph^m 
   + D_{m^*}R_{ij^*kl^*}| \ph^{*m} \right. \non
 && \left.\hs{5} 
    + \1{2} D_n D_m R_{ij^*kl^*}| \ph^n\ph^m 
    + \1{2} D_{n^*}D_{m^*}R_{ij^*kl^*}| 
      \ph^{*m}\ph^{*n} \right. \non
 &&\left.\hs{5}
    + \1{3} (D_{(n^*} D_m R_{ij^*kl^*)_{\rm ah}} 
       + g^{or^*} R_{o(j^*ml^*} R_{in^*k)r^*}
       - g^{or^*} R_{oj^*ml^*} R_{in^*kr^*})|
      \ph^m \ph^{*n} \right] \non 
 && \hs{10} \times 
    \ps^i\ps^k \psb^j\psb^l + O(7).
\eeq
The first two terms are motion terms  
of the bosons and the fermions and 
the others are interaction terms.

We can obtain low-energy theorems of 
scattering amplitudes to $O(p^2)$
(two derivative order) by using the above expression. 
The low-energy scattering amplitudes 
for two bosons can be calculated 
by summing up tree graphs of the fourth order interactions. 
One can obtain the low-energy theorems 
expressed in terms of the curvature tensor of 
a K\"{a}hler manifold, 
since the fourth order term of the bosons is  
the curvature tensor~\cite{HNOO}. 
A calculation of many-body scattering amplitudes 
requires expansions to higher orders.  
One can obtain low-energy theorems expressed in terms of  
the curvature tensor and the covariant derivatives.

%%%%%%%%%%%%%%%

\section*{Acknowledgements} 
We would like to thank Hisao Suzuki for pointing out 
Ref.~\cite{RNC}. 
The work of M.~N. is supported in part 
by JSPS Research Fellowships.

%%%%%%%%%%%%%%%%%%%%%%%%%%%%%
\begin{appendix}
\section{Geometry of K\"{a}hler manifolds}
In this appendix, 
we explain the minimum of K\"{a}hler manifolds. 
(For details see, e.g., Ref.~\cite{Na}.) 
A K\"{a}hler manifold is defined as 
a complex manifold equipped with 
a Hermitian metric and the K\"{a}hler condition 
($d \Omega = 0$ where $\Omega = i g_{ij^*} dz^i \wedge dz^{*j}$). 
As a result of the K\"{a}hler condition, 
the metric can be expressed in terms of a K\"{a}hler potential as  
\beq
 g_{ij^*}(z,z^*) 
  = {\del^2 K(z,z^*) \over \del z^i \del z^{*j}}, 
\eeq
at least in a coordinate patch.\footnote{
To define the metric consistently on the whole 
K\"{a}hler manifold,  
the K\"{a}hler potentials in the union of 
two different patches are related as 
$K\pri(z\pri,z^{\prime *}) 
 = K(z,z^*) + g(z) + g^*(z^*)$, 
where $g$ is a function.  
This is also called a K\"{a}hler transformation, 
like Eq.~(\ref{Kahler-tr.}). 
} 
The connection with mixed indices disappears 
as a result of the compatibility condition 
of the complex structure, $D J = 0$.  
The non-zero connections are given by  
\beq
 {\Gamma^k}_{ij} = g^{kl^*}{\del g_{jl^*} \over \del z^i} 
                 = g^{kl^*} K,_{\,ijl^*}  
\eeq
and their conjugates. 
Derivatives of the metric are
\beq
 && g_{ij^*,k} 
  = {\del g_{ij^*} \over \del z^k} = g_{mj^*}\Gamma^m_{ik}
  = g_{kj^*,i} \defeq \Gamma_{j^*ik},\non
 && g_{ij^*,k^*} 
  = {\del g_{ij^*} \over \del z^{*k}} 
  = g_{im^*}\Gamma^{m^*}_{j^*k^*}
  = g_{ik^*,j^*} \defeq \Gamma_{ij^*k^*}. 
\eeq
Independent components of the curvature tensor are  
\beq
 {R^{i^*}}_{j^*kl^*} 
 = \del_{k}(g^{mi^*} g_{mj^*,l^*})
\eeq
and their conjugates. 
We use the curvature tensor with lower indices: 
\beq
 R_{ij^*kl^*} 
 &\defeq& g_{im^*} {R^{m^*}}_{j^*kl^*}
 = g_{ml^*} {\del \Gamma^m_{ik} \over \del z^{*j}} 
 = {\del^2 g_{kl^*} \over \del z^i \del z^{*j}}
     - g^{mn^*}{\del g_{ml^*} \over \del z^{*j}}
               {\del g_{kn^*} \over \del z^i} \non
 &=& 
  K,_{\,ij^*kl^*}
 - g^{mn^*} K,_{\,mj^*l^*} K,_{\,n^*ik}  \;.\label{cur.}
\eeq
The curvature tensor has the symmetry
\beq
 && R_{ABCD} = - R_{ABDC} = - R_{BACD} = R_{CDAB}, \\ 
 && R_{ij^*kl^*} = R_{kj^*il^*} = R_{il^*kj^*}, 
  \label{sym.of_R2}
\eeq
where the uppercase Roman letters are used for  
both holomorphic and anti-holomorphic indices. 
The second identity is a result of 
the K\"{a}hler condition.

%%%%%%%%%%%%%%%%%%%%%%%%%%%%
\section{A proof of the theorem}
In this section we prove the theorem. 
The starting point is Eq.~(\ref{KNC_exp.}). 
We use the normal coordinates $\omega$, and all 
differentiations are with respect to $\omega$ in this section. 
Before giving a proof of the theorem, we prove a lemma. 

We denote (a set of) the K\"{a}hler potential 
differentiated at most $n$ times as $K,_{(n)}$. 
($K,_{(n)} \subset K,_{(n+1)}$.) 
For example, $K,_{\,ijk^*} \in K,_{(3)}$. 
Note that, 
although all terms with $(n,1)$ and $(1,n)$ indices, 
$K,_{\,i_1\cdots i_n j_1^*}$ and 
$K,_{\,ij_1^*\cdots j_n^*}$,  
vanish at the origin ($\omega=0$),  
$K,_{\,i_1\cdots i_n j_1^*}| = 
K,_{\,ij_1^*\cdots j_n^*}|= 0$, 
they do not disappear at an arbitrary value of $\omega$ in general. 
(For example, the connection (\ref{connection_exp}) vanishes 
only at the origin.)  
The theorem states that the remaining $K,_{(N)}$ 
become covariant tensors {\it at the origin}.  

If we fix the ordering of the holomorphic and 
anti-holomorphic covariant derivatives, 
there is a one-to-one correspondence between 
the $(M,N-M)$ tensor  
\beq
 {R^{(N)}}_{j_1^*\cdots j_{N-M-2}^* i_1 \cdots i_{M-2}ij^*kl^*}
 \defeq 
 D_{j_1^*}\cdots D_{j_{N-M-2}^*}D_{i_1}\cdots D_{i_{M-2}}
   R_{ij^*kl^*} \nonumber 
\eeq  
and the coordinate derivative of the K\"{a}hler potential 
$K,_{\,j_1^*\cdots j_{N-M-2}^* i_1 \cdots i_{M-2} ij^*kl^*}
\in K,_{(N)}$.  
(For example, see Eqs.~(\ref{cur.}), (\ref{(3,2)}) and 
(\ref{D*DR}) for the first few orders.)  
By generalizing these equations, 
we obtain the following lemma giving 
a relation between the covariant $(M,N-M)$ tensor 
and the coordinate derivative of the K\"{a}hler potential. 

{\bf Lemma.} 
The curvature tensor 
covariantly differentiated $(M-2,N-M-2)$ times 
can be written as ($N \geq 4$, $2\leq M\leq N-2$)
\beq
 && \hs{5} 
  {R^{(N)}}_{j_1^*\cdots j_{N-M-2}^* i_1 \cdots i_{M-2}ij^*kl^*}\non
 &&\in K,_{\,j_1^*\cdots j_{N-M-2}^* i_1 \cdots i_{M-2} ij^*kl^*}
 + \sum_{\alpha=1}^{N-3} (-1)^{\alpha}(g^{-1})^{\alpha} 
   \underline{K,_{(N-1)} \cdots K,_{(N-1)}}. \label{asp} \\ 
 && \hs{80} \mbox{($\alpha+1$)-times} \nonumber
\eeq
The first term in the second line is an element of $K,_{(N)}$,  
and all terms have $M$ holomorphic and $(N-M)$ 
anti-holomorphic indices.
The $g^{-1}$ in the second term 
denotes the inverse metric $g^{ij^*}$. 
Each $g^{-1}$ contracts indices of two different 
$K,_{(N-1)}$'s.\\  
(Proof) 
We use mathematical induction for the proof.
\begin{enumerate}
\item 
First, we show that Eq.~(\ref{asp}) holds for $N=4$:\\  
${R^{(4)}}_{ij^*kl^*} 
 = R_{ij^*kl^*} = K,_{\,ij^*kl^*} 
 - g^{mn^*} K,_{\,mj^*l^*} K,_{\,ikn^*}$ is 
in the form of Eq.~(\ref{asp}).
(This is trivial when $N$ is less than $4$.)

\item
Second, we show Eq.~(\ref{asp}) for ($N+1$)-th order,  
assuming that it holds at $N$-th order.   
One of elements at $(N+1)$-th orders is 
\beq
 &&\hs{5} 
   {R^{(N+1)}}_{j_1^*\cdots j_{N-M-1}^*i_1\cdots i_{M-2}ij^*kl^*}
   = D_{j_{N-M-1}^*} 
   {R^{(N)}}_{j_1^*\cdots j_{N-M-2}^* i_1 \cdots i_{M-2}ij^*kl^*}\non
 &&=\del_{j_{N-M-1}^*} 
    {R^{(N)}}_{j_1^*\cdots j_{N-M-2}^*i_1\cdots i_{M-2} ij^*kl^*}\non
 && 
   - \sum_{a=1}^{N-M-2} {\Gamma^{m^*}}_{j_{N-M-1}^* j_a^*} 
  {R^{(N)}}_{j_1^*\cdots j_{a-1}^* m^* j_{a+1}^* \cdots 
       j_{N-M-2}^* i_1 \cdots i_{M-2} ij^*kl^*} \non
 && 
   - {\Gamma^{m^*}}_{j_{N-M+1}^* j^*} 
      {R^{(N)}}_{j_1^*\cdots j_{N-M-2}^*i_1\cdots i_{M-2}im^*kl^*}\non 
 && 
   - {\Gamma^{m^*}}_{j_{N-M+1}^* l^*} 
      {R^{(N)}}_{j_1^*\cdots j_{N-M-2}^*i_1\cdots i_{M-2}ij^*km^*}.  
  \label{pr1}
\eeq
Since ${\Gamma^i}_{jk} = g^{il^*} K,_{jkl^*} \in g^{-1} K,_{(3)}$, 
the last three terms are of the form\\ 
$- \sum g^{-1} K,_{(3)} {R^{(N)}}$. 
Moreover, this can be rewritten in the form 
\beq
 && \sum_{\alpha=1}^{N-2} (-1)^{\alpha}(g^{-1})^{\alpha} 
   \underline{K,_{(N)} \cdots K,_{(N)}} \;\;, \label{pr2} \\
 && \hs{30} \mbox{($\alpha+1$)-times} \nonumber
\eeq
since $K,_{(3)} \subset K,_{(N-1)} \subset K,_{(N)}$. 
The first term on the right-hand side of 
Eq.~(\ref{pr1}) has the form 
\beq
 && \del_{j_{N-M-1}^*} 
     K,_{\,j_1^*\cdots j_{N-M-2}^* i_1 \cdots i_{M-2}ij^*kl^*}\non
 && 
   + \sum_{\alpha=1}^{N-3} (-1)^{\alpha}
  \left[- (g^{-1})^{\alpha+1} K,_{(3)} 
   \underline{K,_{(N-1)} \cdots K,_{(N-1)}} 
       + (g^{-1})^{\alpha} 
   \underline{K,_{(N)} \cdots K,_{(N)}} \right] ,\non
 && \hs{55} \mbox{($\alpha+1$)-times}  
    \hs{25} \mbox{($\alpha+1$)-times} \non
\eeq
where we have used $\del g^{-1} 
= - (g^{-1})^2 K,_{(3)}$ from Eq.~(\ref{del_g}). 
The second term on the right-hand side 
is also of the form of Eq.~(\ref{pr2}).
We thus have proved 
\beq
 && \hs{5}
  {R^{(N+1)}}_{j_1^*\cdots j_{N-M-1}^*i_1\cdots i_{M-2}ij^*kl^*}\non
 && \in K,_{\,j_1^*\cdots j_{N-M-1}^* i_1 \cdots i_{M-2} ij^*kl^*}
 + \sum_{\alpha=1}^{N-2} (-1)^{\alpha}(g^{-1})^{\alpha} 
   \underline{K,_{(N)} \cdots K,_{(N)}} . \label{proof}\\
 && \hs{80} \mbox{($\alpha+1$)-times} \nonumber
\eeq
The right-hand side has the same form as Eq.~(\ref{asp}), 
where the first term is an element of $K,_{(N+1)}$.

We also have to show the lemma for another element of 
$(N+1)$-th order,  
${R^{(N+1)}}_{j_1^*\cdots j_{N-M-2}^*i_1
\cdots i_{M-1}ij^*kl^*}$.  
The difference between this standard ordering 
and the non-standard ordering 
$D_{j_{M-1}} {R^{(N)}}_{j_1^*\cdots j_{N-M-2}^* i_1 
\cdots i_{M-2}ij^*kl^*}$ can be written 
as products of the curvature tensor and 
terms from $R^{(4)}$ to $R^{(N)}$ 
as a result of Eq.~(\ref{com.D-D}). 
Since we can show that the latter can be written in the form of 
Eq.~(\ref{asp}), as in the same manner above, 
the former can be also written in the form of 
Eq.~(\ref{asp}). 

\item From 1 and 2, 
we prove (\ref{asp}) for any $N (\geq 4)$. 
(Q.E.D.)
\end{enumerate}

Now we are ready to give a proof of the theorem.
We would like to show that all coefficients in the expansion 
(\ref{KNC_exp.}) $K,_{\, i_1 \cdots i_M j_1^* \cdots j_{N-M}^*}|$
{\it evaluated at the origin} $\omega=0$ can be expressed by covariant 
quantities {\it at the origin}. 
Now, we have the relation Eq.~(\ref{asp}) 
between these coefficients and covariant tensors.
The left-hand side is, of course, covariant. 
Our task is to show that each term 
on the right-hand side evaluated at the origin is also covariant. 
\\  
(Proof of the theorem.) 
We again use mathematical induction for the proof. 
Note that $K,_{(3)}|$ vanishes, by our choice of the K\"{a}hler 
normal coordinates.
\begin{enumerate}
\item 
By the equation $R_{ij^*kl^*} = K,_{\,ij^*kl^*} 
 - g^{mn^*} K,_{\,mj^*l^*} K,_{\,ikn^*}$, 
we can show that 
$K,_{(4)}| \ni  K,_{\,ij^*kl^*}| = R_{ij^*kl^*}|$ is covariant.  
($K,_{(2)} \ni g_{ij^*}$ is covariant at any $\omega$.)

\item 
We assume that all terms of order less than $(N+1)$,  
$K,_{(4)} \cdots K,_{(N)}$, 
are covariant {\it at the origin}.  
(Terms with $(n,1)$ and $(1,n)$ indices disappear at the origin.)
Then $K,_{(N+1)}|$ is also expressed in terms of 
covariant quantities, by Eq.~(\ref{proof}), 
since all other terms are covariant, by assumption. 

\item From 1 and 2, all of the coefficients in 
Eq.~(\ref{KNC_exp.}) are covariant.  
(Q.E.D.)
\end{enumerate}

\end{appendix}

%%%%%%%%%%%%%%%%%%%%%%%%%%%%%%%%%%%%%%%%%%%%%%%%%%%%%%%%%%%%%%%%%%%%%%%%%%%

\end{document}